# Efficient Passivation of Surface Defects by Lewis Base in Lead-free Tin-based Perovskite Solar Cells

Hejin Yan[1], Bowen Wang[1], Xuefei Yan[1], Qiye Guan[1], Hongfei Chen[1], Zheng Shu[1], Dawei Wen[2], Yongqing Cai[1*]

[1]Joint Key Laboratory of the Ministry of Education, Institute of Applied Physics and Materials Engineering, University of Macau, Taipa, Macau, China

[2]School of Applied Physics and Materials, Wuyi University, Jiangmen, China

\* Corresponding authors

E-mail: yongqingcai@um.edu.mo

**ABSTRACT:** Lead-free tin-based perovskites are highly appealing for the next generation of solar cells due to their intriguing optoelectronic properties. However, the tendency of $Sn^{2+}$ oxidation to $Sn^{4+}$ in the tin-based perovskites induces serious film degradation and performance deterioration. Herein, we demonstrate, through the density functional theory based first-principle calculations in a surface slab model, that the surface defects of the Sn-based perovskite FASnI$_3$ (FA = NH$_2$CHNH$_2^+$) could be effectively passivated by the Lewis base molecules. The passivation performance of Lewis base molecules in tin-based perovskite is tightly correlated with their molecular hardness. We reveal that the degree of hardness of Lewis adsorbate governs the stabilization via dual effects: first, changing the stubborn spatial distribution of tin vacancy (V$_{Sn}$) by triggering charge redistribution; second, saturating the dangling states while simultaneously reducing the amounts of deep band gap states. Specifically, the hard Lewis base molecules like edamine (N-donor group) and Isatin-Cl (Cl-donor group) would show a better healing effect than other candidates on the defects-contained tin-based perovskite surface with a somehow hard Lewis acid nature. Our research provides a general strategy for additive engineering and fabricating stable and high-efficiency lead-free Sn-based perovskite solar cells.

## 1. Introduction

Perovskite solar cells (PSCs) have been under the spotlight due to their superior optoelectronic properties and solution processability for the next generation of photovoltaic technology.[1, 2] In the past decade, the power conversion efficiency (PCE) of lead-based PSCs boost rapidly from 3.8% to 25.7%, comparable with the commercial silicon solar cell.[3] However, the toxicity of lead and potential environmental hazard hinder their practical application.[4] As an alternative, a series of lead-free perovskite materials, including Sn, Bi, and Ge-based perovskites have been widely explored.[5-8] In particular, tin-based alternatives resemble their lead-contained counterpart, as the neighboring Sn and Pb elements in the periodic table share a similar valence structure.[9] The alike electronic configurations and atomic radius between Pb and Sn stabilize the octahedral inorganic $ABX_3$ (A = organic or inorganic cation, B = metal ion, X = halogen) frameworks, leading to comparable electronic structure and carriers dynamics.[10] However, so far the highest PCE of 14.81% achieved in $FASnI_3$ PSCs is generally inferior to that of lead-based PSCs.[11] The main reason for this inferiority is the easy oxidation of $Sn^{2+}$ to $Sn^{4+}$ in the tin-based perovskite, leading to a degradation of performance.[12, 13]

From a chemical perspective, lone pair electrons in $Sn^{2+}$ have a stronger chemical activity compared with that of $Pb^{2+}$ due to the absence of the lanthanide shrinkage effect.[14] The $Sn^{2+}$ is prone to lose two 5$s$ electrons transferred to $Sn^{4+}$, accounting for the oxidation of tin-based perovskite. This oxidation process further accelerates the decomposition of perovskite film[15, 16] akin to the process:

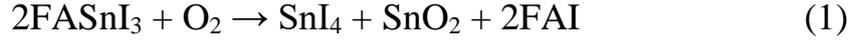

$$2FASnI_3 + O_2 \rightarrow SnI_4 + SnO_2 + 2FAI \qquad (1)$$

A recent work also proved the strong coupling between film oxidation and tin vacancy ($V_{Sn}$) defect formation, which is the dominant type of defect for both the Sn-rich and I-rich situations in $FASnI_3$.[17]

Moreover, the energy level of Sn-5s orbital is higher than that of Pb-6s orbital, impelling both the defect formation and tin oxidation.[18] The lowered redox potential of $Sn^{2+}/Sn^{4+}$ at +0.15 eV makes $Sn^{2+}$ easily oxidized to $Sn^{4+}$.[19, 20] This is associated with a higher energetically lying antibonding orbital and thus weaker Sn-I bonds than the Pb-contained compounds. The breaking of Sn-I bond also facilitates the formation of $V_{Sn}$ defects [17] via:

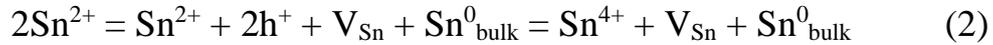

$$2Sn^{2+} = Sn^{2+} + 2h^+ + V_{Sn} + Sn^0_{bulk} = Sn^{4+} + V_{Sn} + Sn^0_{bulk} \qquad (2)$$

here, the first equation from the left means the formation of each $V_{Sn}$ is accompanied with the release of two holes and a neutral interstitial Sn atom. The second equation suggests that those two holes can be further trapped by a $Sn^{2+}$ which is oxidized into $Sn^{4+}$. Therefore, as dictated by equation (1) and (2), both the Lewis acid type $Sn^{2+}$ oxidation and the generation of surface $V_{Sn}$ defects occur simultaneously. Accumulation of such defects is responsible for the degraded performance and structures in tin-based PSCs, usually associated with defects induced loss of open voltage ($V_{oc}$), high background hole density, and blue shift in absorption edge.[21, 22]

To effectively mitigate the defects and associated effects shown above, reductive organic agents are introduced during the fabrication of tin-based PSCs.[23, 24] These additives act as the Lewis base to stabilize the defects-contained perovskite (be hard Lewis acid like) by supplying extra charge.[25] For example, Kamarudin et al. used the edamine to passivate surface dangling bonds of FASnI$_3$ which act as non-radiative recombination centers, a long diffusion distance of up to 300 nm and high efficiency of 10.18% were obtained after passivation.[26] On the other hand, the inclusion of organic additives with different functional groups lead to significantly diverse performance of defect passivation.[26] Gu et al. proposed a novel 2,3-diaminopropionic acid monohydrochloride (2,3-DAPAC) molecule into FASnI$_3$ perovskite film.[27] The nuclear magnetic resonance (NMR) spectroscopy indicated there is a coordinate interaction between -NH$_2$ and Sn$^{2+}$, together an electrostatic attraction between -COO$^-$ and Sn$^{2+}$. This strategy significantly reduced V$_{Sn}$ defect density and suppressed tin oxidation reflected by the increased $V_{oc}$ and reduced dark current density.[27] Similarly, Park et al. investigated the interaction between the reductive surface ligand and perovskite film through adopting different binding groups, among which the carboxyl-contained *p*-Toluic acid (*p*-TA) molecules showed the best improvement of efficiency, and significant shifts of ionization energy and electron affinity by adsorption of additives were observed.[28] Ethanedithiol (EDT) molecule with a hydrophobic nature has also been extensively used in the PSCs due to its good passivating ability.[29, 30] Xiong et al. used a kind of organic dye named 5-chloroisatin (Isatin-Cl) to passivate the uncoordinated defects and eliminate the p-type doping within the film.[31] Some representative Lewis base additive usages, regardless of the final PCE, are listed in Table. S1 for comparison. These experiments suggest the interaction between the organic molecules and defects bearing surface is critical for the modulated efficiency of tin-based PSCs. Nevertheless, most of the reported studies were still based on the trial-and-error approach, and passivating mechanism of the functional Lewis base additives to surface defects-contained tin-based PSCs is still not understood.

In this study we systematically investigated the healing effect of multiple functional passivators, including the edamine (N-donor), *p*-TA (O-donor), EDT (S-donor), and Isatin-Cl (Cl-donor), to the defects-contained FASnI$_3$ perovskite surface. We conduct the state-of-the-art first-principle calculations with the density functional theory (DFT) method, which reveal the passivation mechanism of Lewis base molecules by comparing charge transfer, broadened band gap, and charge delocalization associated with the defect. The effects of molecular concentration and other surface vacancies, e.g., iodine vacancy, are neglected to reduce the complexity of the models. In addition, to be more focused and make a smoother illustration, we narrow the analysis down to the healing effect of the passivated molecule without discussing any likeliness of suppressed formation of defects of molecule-protected surface, albeit the latter would be another positive effect of surface decoration.[32] Interestingly, the interaction between the passivating ligand and surface defects in tin-based perovskite is strongly correlated to the molecular hardness. The hard Lewis base molecules can effectively compensate the holes states associated with the $V_{Sn}$ and other undercoordinated Sn atoms. These results shed light on the atomic-scale mechanisms of promoting the efficiencies of PSCs via charge-compensation engineering.

## 2. Computational details

The calculations using density functional theory (DFT) were performed based on the projector augmented wave (PAW) potential method implemented in the Vienna *ab initio* simulation package (VASP).[33, 34] The generalized gradient approximation (GGA) of the Perdew-Burke-Ernzerh function (PBE) was employed to describe the exchange-correlation potential.[35] By setting a 500 eV cutoff energy on the wave function, a 2×2×1 and another denser 3×3×1 Monkhorst-Pack k-point mesh were applied for the surface model optimization and the electronic properties calculation, respectively. The van der Waals interactions were described by the DFT-D3 method.[36] The convergence criteria were set to $10^{-3}$ and $10^{-5}$ eV for both the atomic force and energy. The spin-orbital coupling was not considered due to the non-significant relativity effect in the tin element and charge transfer process. The Bader charges were calculated by using the method of Hen Kelman et al.[37] The optimized cell structure parameters of the tetragonal phase $FASnI_3$ unit cell are a = b = 8.859 Å, and c = 12.651 Å, which show good agreement with the experiment results.[38, 39] The surface models are constructed based on this bulk $FASnI_3$ structure with the bottom atomic layer be fixed. A vacuum layer with a thickness of 15 Å was added along the z-direction to avoid the impact of image atoms.

## 3. Results and Discussion

### 3.1 Localized state of pristine surface of $FASnI_3$:

Figure 1a and b show the optimized $FASnI_3$ surface structures without and with $V_{Sn}$ defect, respectively. The presence of $V_{Sn}$ breaks the skeleton connectivity and induces the distortion of the $SnI_5$ octahedron around the defect. The removal of one surface Sn atom leads to five dangling iodine atoms which are relaxed outwardly from the $V_{Sn}$. There is an enhanced hybridization and chemical bonding of these dangling iodine atoms with their Sn neighbors as indicative of the shortened Sn-I bond length decreasing from 3.15 Å for the pristine surface to 2.92 Å. The electronic structure of the defects is an underlying factor that governs the polarity of carriers and response of photo excitation. The spatial distribution of the defective states of $V_{Sn}$ is plotted in Fig. 1b. Surprisingly, the defective states do not locate at the dangling I atoms as normal vacancy states do. Instead, such a localized state is found at the neighboring Sn atom toward the bulk interior and penetrates several atom layers, indicating a strong redox ability of $V_{Sn}$ for producing $Sn^{2+}/Sn^{4+}$ oxidation. Thus, the presence of $V_{Sn}$ defect tends to cause the oxidation of adjacent Sn atoms and the surface structure unstable. This self-doping mechanism is responsible for the degraded performance and structural instability of the non-stoichiometric tin-based perovskite.[19, 40] To this end, we aim to passivate such surface states by introducing the Lewis base molecules.

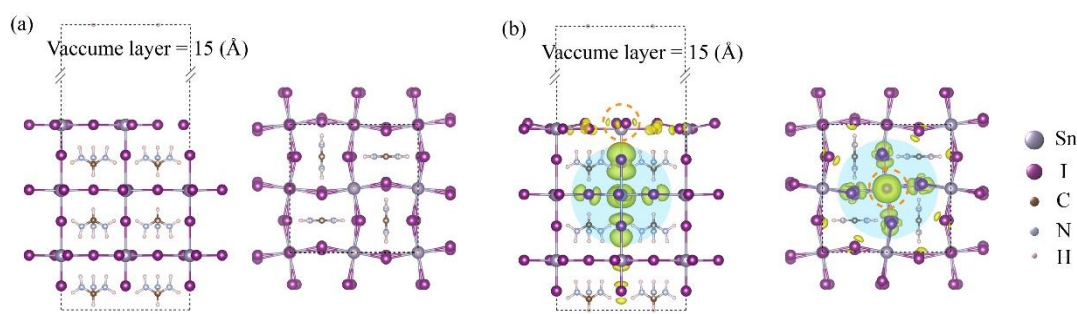

**Figure 1.** The side and top views of the optimized tin perovskite surface with (a) and without (b) the $V_{Sn}$. The position of $V_{Sn}$ and spatial location of its defect level are marked by the dashed red circle and nattier blue zone, respectively. The iso-surface value of charge density is $0.001 e/Bohr^3$.

**3.2 Energetics of adsorption of Lewis base molecules:**

Herein we choose several typical Lewis base molecules including edamine, *p*-TA, EDT, and Isatin-Cl molecules. We consider the Sn deficient FASnI$_3$ (001) surface where a V$_{Sn}$ is created at the top surface plane (Fig. 1b). As expected, such molecules are less likely to be adsorbed at the V$_{Sn}$ core where weak van der Waals interactions occur between the molecule and dangling iodine groups. The Lewis base molecules passivate the V$_{Sn}$ through an indirect way, the molecules bonding with the exposed Sn atoms surrounding the V$_{Sn}$ core rather than the V$_{Sn}$ core itself. The edamine molecule interacts with the exposed Sn atoms by forming a covalent Sn-N bond with a length of 2.60 Å; the passivation of *p*-TA through the Sn-O bond with the shortest bond length of 2.05 Å; the EDT adsorbed through the Sn-S bond with a bond length of 2.46 Å; the Isatin-Cl anchors through Sn-Cl bond with the longest bond length of 2.92 Å (Fig. S2). Meanwhile, the octahedron distortion in the upper layer SnI$_5$, as shown in Table. 1, is released for most cases except the EDT passivation. To explain this surface interaction, the surface adsorption energy and the charge density difference are calculated. The adsorption energy ($E_{ads}$) between the adsorption molecules and perovskite substrate can be written as:

$$E_{ads} = E_{total} - E_{molecule} - E_{perovskite} \qquad (3)$$

where the $E_{total}$ is the total energy of the surface model, $E_{molecule}$ and $E_{perovskite}$ are the energy of the isolated molecule and perovskite surface, respectively.[48] The values of $E_{ads}$, listed in Table. 1, vary from strong binding to weak adsorption as the sequence: p-TA (-5.47 eV) > EDT (-4.41 eV) > edamine (-1.95 eV) > Isatin-Cl (-0.44 eV). Thus, molecule adsorption is energetically favourable for all these surface passivation cases. Furthermore, the largest $E_{ads}$ of p-TA indicates the strongest adsorption between the carboxyl group and the perovskite surface. These adsorbed organic ligands also serve as moisture and oxygen scavengers for further protection of the inner perovskite layer which is known as the protective coating method.[32, 41]

**Table 1.** Adsorption of Lewis base molecules above Sn deficient FASnI$_3$ surface: the average bond length ($d_1$) in surface SnI$_5$ unit, distance ($d_2$) between the Sn and molecular anchoring atoms, adsorption energy ($E_{ads}$), and transferred electrons ($e_{transfer}$) of the passivation molecules on the perovskite surface. Note a positive value of $e_{transfer}$ means a loss of electrons from Lewis base molecules to perovskite.

| Model | $d_1$/Å | $d_2$/Å | $E_{ads}$/eV | $e_{transfer}$/10$^{-3}$e |
|---|---|---|---|---|
| perovskite without V$_{sn}$ | 3.15 | — | — | — |
| perovskite with V$_{sn}$ | 2.90 | — | — | — |
| edamine/perovskite | 2.96 | 2.60 | -1.95 | 1.64 |
| Isatin-Cl/perovskite | 2.92 | 2.92 | -0.44 | 0.40 |
| *p*-TA/perovskite | 2.92 | 2.05 | -5.47 | -4.42 |
| EDT/perovskite | 2.89 | 2.46 | -4.41 | -3.06 |

**3.3 Charge transfer and passivation of localized defective states**:

Similarly, the charge transfer between the molecules and perovskite surface can be obtained according to the formula:

$$\rho_{transfer} = \rho_{total} - \rho_{molecule} - \rho_{perovskite} \qquad (4)$$

where the $\rho_{transfer}$ is the charge density of the surface model, $\rho_{molecule}$ and $\rho_{perovskite}$ are the charge density of the isolated molecule and perovskite surface, respectively.[42] In Fig. 2, the charge gain and loss regions are marked by red and green, respectively. The quantitative estimation of the amount of charge transfer is listed in Table. 1 through spatial integration of the $\rho_{transfer}$. We find that the edamine, and Isatin-Cl molecules transfer their electrons to passivate the surface defect, where a charge accumulation region is formed in the perovskite side. The largest amount of charge transfer in the *p*-TA/perovskite system (Fig. 2c) is consistent with its largest $E_{ads}$, while the Isatin-Cl (Fig. 2b) has the smallest charge transfer, as the Cl atom has a relatively weak interaction with the perovskite surface. From our calculation, the Lewis base additives mainly interact with the undercoordinated Sn atoms to maintain the octahedral backbone structure of the perovskite surface. This is consistent with previous experimental results, that Lewis base additives mediate the defects-contained perovskite surface.[43]

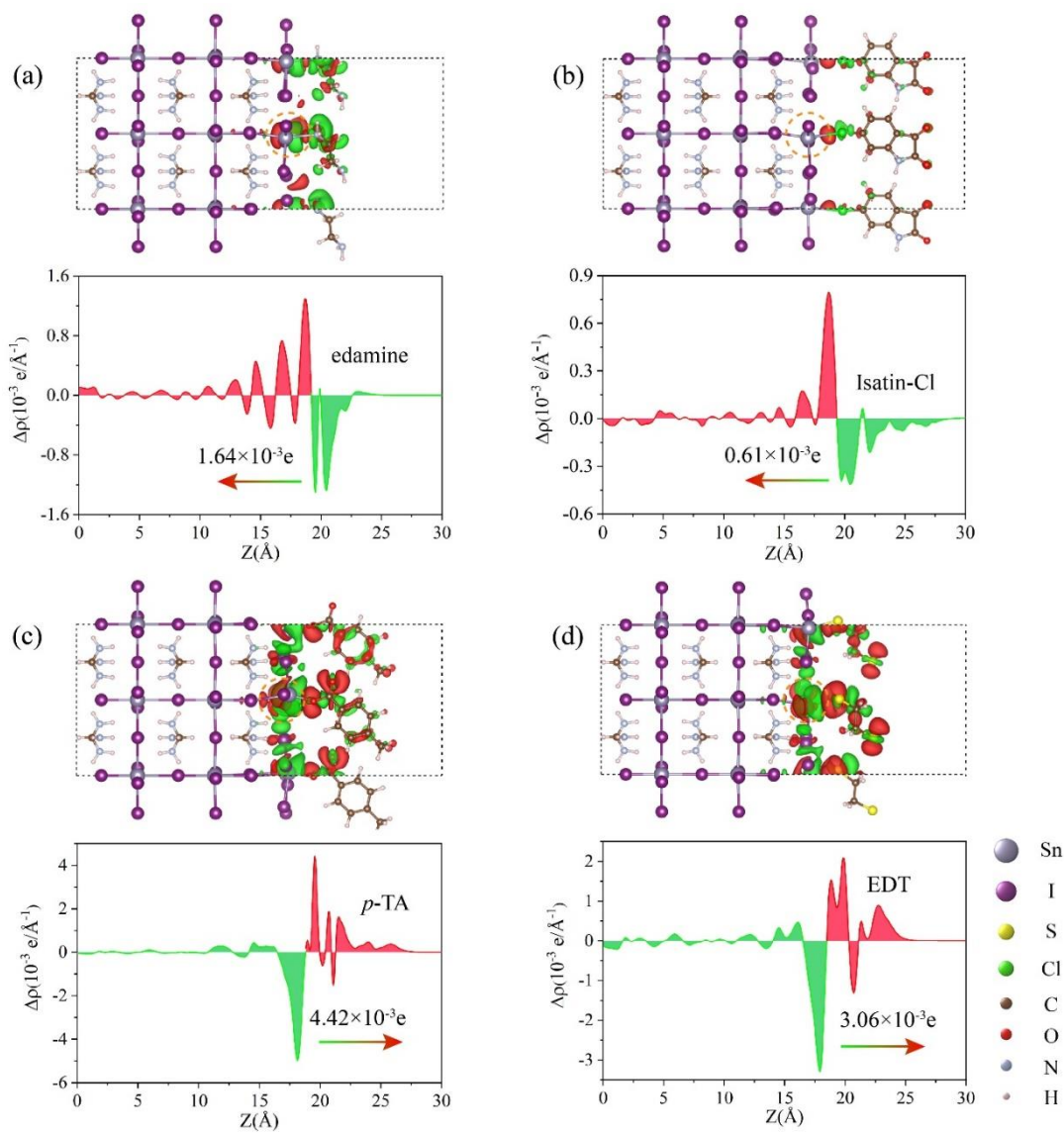

**Figure 2.** The charge density difference and in-plan average for the (a) edamine/perovskite, (b) Isatin-Cl/perovskite, (c) *p*-TA/perovskite, and (d) EDT/perovskite, respectively. The charge accumulation region is marked in orange with the charge depletion region marked in green.

Theoretically, the Lewis base molecules are prone to transfer electrons to the exposed tin atoms, the transferred excess electrons accumulate at the perovskite side and thus stabilizing the tin perovskite surface by suppressing the oxidation. Here, the planar average charge density difference curve along the Z direction is calculated which allows a quantitative estimation of the charge accumulation and depletion across the entire interface. Interestingly, as shown in Fig. 3, the trend and amounts of surface charge transfer correlate with the hardness ($\eta$) of the Lewis base. The chemical hardness of the molecules is an index of resistance to change their electronic configuration.[44] The transferred electrons of edamine ($\eta_{edamine}$ = 3.28 eV) are larger than the Isatin-Cl ($\eta_{Isatin-Cl}$ = 2.42 eV). Thus, the edamine has a stronger stabilizing ability than Isatin-Cl. While the *p*-TA ($\eta_{p\text{-}TA}$ = 1.98 eV) and EDT (with $\eta_{EDT}$ = 1.82 eV) cause an inverse charge transfer which is unfavorable for surface defect passivation.[45] The altered amount of charge transfer between the molecules and perovskite surface can be contributed to the different interactions at the adsorbate-perovskite interface. Besides, the average Sn-I bond length also has a strong correlation with the chemical hardness value. The longer the Sn-I bond length, the larger the electron transfer value, which indicates the passivation of the surface and suppression of tin oxidation. The importance of charge compensation for $V_{Sn}$ is realized until recently, through which the background hole can be largely neutralized, hence the prolonged charge carrier lifetime and promoted mobility.[46] Also, this compensation can release the $V_{Sn}$ induced lattice strain.[47, 48]

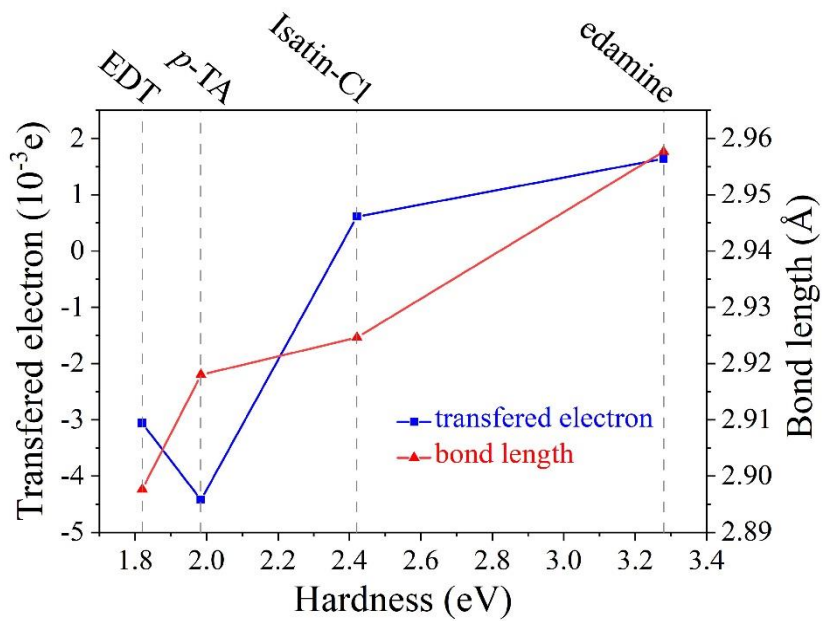

**Figure 3.** The Sn-I bond length and charge transfer as a function of the molecular hardness.

We next performed the electron localization function (ELF) analysis to illustrate the electronic coupling between the organic and perovskite surface. The ELF value ranges from 0 to 1, which represents the electronic states from a strong localization state to a low-density state.[49] As shown in Fig. 4, the Sn-N and Sn-O bonds of the edamine/perovskite and *p*-TA/perovskite systems have an obvious covalent character as the ELF value at the bond close to 0.5. On the contrary, the Sn-S bond in the EDT/perovskite surface has a strong ionic bond character. The Sn-I bond of the Isatin-Cl/perovskite system also has a covalent character, but its strength is weaker than that of the Sn-I bond. This is also consistent with the small $E_{ads}$ and the low charge transfer between the Isatin-Cl and perovskite. Obviously, the functional ligands have stable interaction with uncoordinated tin perovskite, which is critical for forming a dense protective layer to prevent the penetration of moisture and oxygen.[9, 50]

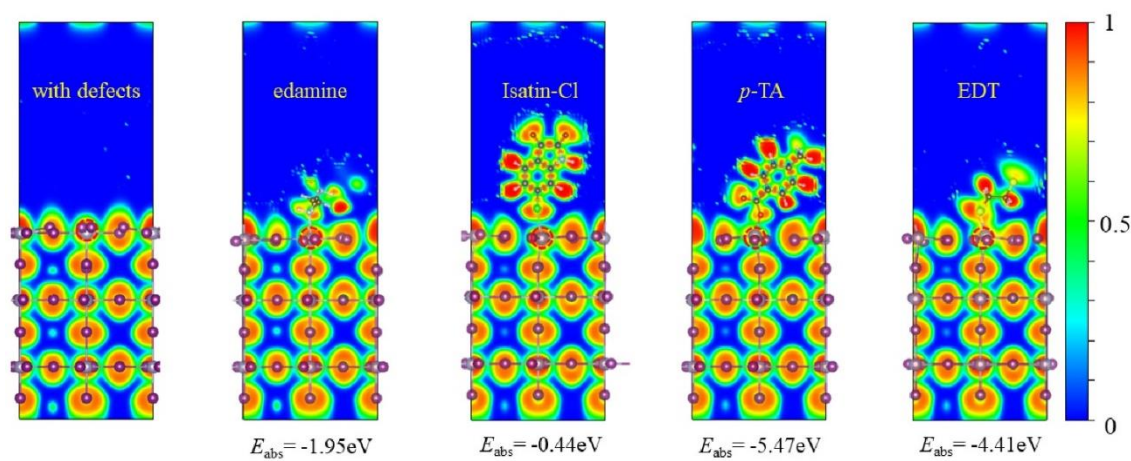

**Figure 4.** The ELF value of the Lewis base molecule passivated surfaces.

The density of states (DOS) of the defects-contained surface is shown in Fig. S1, where the band gap has been submerged in the multiple surface states. These deep levels, including the surface $V_{Sn}$ defects, act as electron traps and recombination centers in the tin-based PSCs. Thus, the reductive molecule additives are designed to suppress these defects. The enlarged band gap and reduced metallic density near the Fermi level in Fig. 5 can be attributed to the effective healing of defect healing, where the surface states have been partially compensated by the surface molecule. According to the partial density of states (PDOS) value, the $V_{Sn}$ serves as the nonradiative recombination center lying in the deep level, while the unsaturated tin atoms exposed as the surface induce considerable shallow levels. Previous calculations and experiments highlight the background hole states caused by $V_{Sn}$ in tin perovskite, also the $V_{Sn}$ defects at grain boundaries and surface are regarded as an important source for $V_{oc}$ loss.[17, 22] Surface passivation can eliminate the number of shallow defects and shift the positions of deep defects, and the band gap increases accordingly. In addition, after surface decoration, the Bader charges of outmost tin atoms, are more similar with the $Sn^{2+}$ state in the bulk phase $FASnI_3$ rather than the $Sn^{4+}$ state in the bulk phase $FA_2SnI_6$. The averaged Bader values are listed in Table. S2. Therefore, the charged defects on the perovskite surface have been stabilized through interacting with the functional ligands.

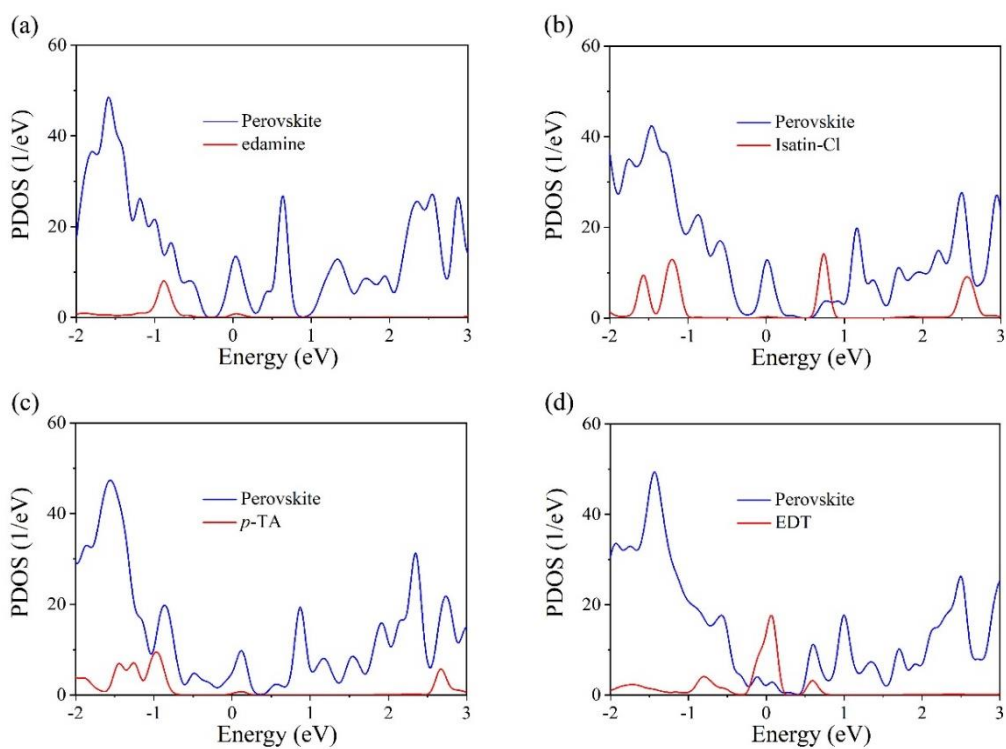

**Figure 5.** The PDOS of the (a) edamine/perovskite, (b) Isatin-Cl/perovskite, (c) *p*-TA/perovskite, and (d) EDT/perovskite interfaces, respectively.

**3.4 Mechanism and effectiveness of Lewis base molecules:**

We next discuss the effectiveness of the Lewis base molecules on passivation of the defective states of $V_{Sn}$. As discussed above, each $V_{Sn}$ is likely to induce excess holes which cause the oxidation of the nearby $Sn^{2+}$ to $Sn^{4+}$, thus the healing effect to these vacancies induced background charge is so-called charge compensation. Upon the introduction of the external dopant such as the Lewis base molecules shown here, the additional charged center of these adsorbates would disturb the spatial distribution of the defective states and achieve passivation. As shown in Fig. 6, however, the effect of the passivation seems to vary with molecules. Upon the adsorption of EDT, the $V_{Sn}$ associated defective state is still located at the Sn atom beneath the $V_{Sn}$ site, which suggests a remarkable barrier for the polaron hopping around these two Lewis adsorbates. In contrast, interestingly, the uptake of edamine, *p*-TA, and Isatin-Cl above the surface both induce an upward shift of the defective state toward the surface adsorbate, implying a significant electronic coupling between the $V_{Sn}$ and such Lewis complex. Our calculations reveal that edamine and Isatin-Cl are more effective to compensate the $V_{Sn}$ associated hole states compared with the intrinsic $Sn^{2+}/Sn^{4+}$ transition. According to the classical hard-soft-acid-base theory (HSAB), edamine and Isatin-Cl are regarded as hard Lewis base, and thus they are more prone to interact with the $V_{Sn}$ contained perovskite surface which is hard Lewis acid.[51, 52] The Lewis base with large hardness transfers the electron to the perovskite layer and passivate the surface defects, accordingly an enlarged band gap is achieved for edamine and Isatin-Cl passivated structures. Although the defects can't be entirely neutralized by the Lewis base, rational designed molecular dopant alters the electronic screening and accommodate excess holes thus effectively reducing the tin deficient surface. Pascual et al. systematically compared the function of SnX (X = F, Cl, Br, and I) additive to the reducing $Sn^{4+}$ content in the precursor solution and the final perovskite film. They also suggest that the harder Lewis bases are the more suitable additives for processing the tin-based perovskites.[53]

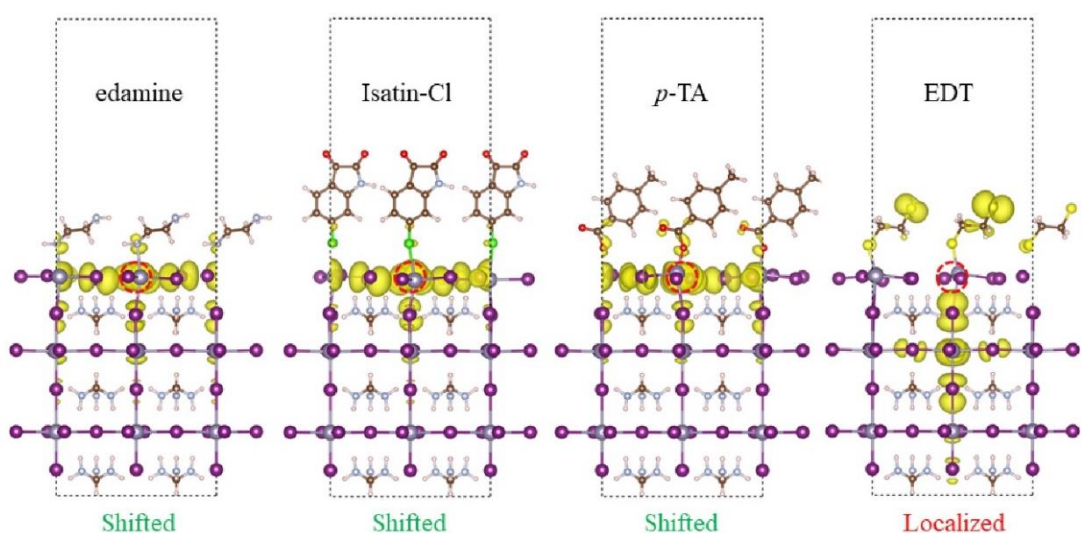

**Figure 6.** Lewis base molecule modulated spatial distribution of $V_{Sn}$ defect states. The charge density of the surface defects is plotted with the iso-surface value of $0.001e/Bohr^3$.

Here our work shows that different functional groups in the molecules may affect the adsorption and charge transfer. For instance, the soft Lewis base *p*-TA and EDT passivate the surface by strong interaction with the uncoordinated Sn atoms. But an inverse charge transfer, from perovskite to molecules, makes little contribution to background hole density neutralization. Comparingly, the hard Lewis base edamine and Isatin-Cl stabilize surface defects through charge compensation. Their moderate interaction intensity and charge transfer are also favorable for the starching of surface Sn-I bonds, hence the reduced lattice strain. To further prove our idea of the passivation of defects through Lewis base, we also examined the adsorption of soft Lewis base molecules above the FASnI$_3$ surface. Herein the benzene sulfonic acid (BSA) as a soft Lewis base was considered. As shown in Fig. S2, the passivation effect of the BSA is close to that of the sulfur-contained EDT and carboxyl-contained *p*-TA. Thus, the Lewis base mainly influences the perovskite surface through two aspects: Firstly, the functional ligand coordinates with the undersaturated tin atoms, the following charge compensation reduce the SnI$_5$ distortion and avoid further tin oxidation (Fig. 7a); Secondly, surface passivation can eliminate the amounts of shallow defects (unsaturated tin atoms) and shift the deep defects ($V_{Sn}$) to shallow defects (Fig. 7b).

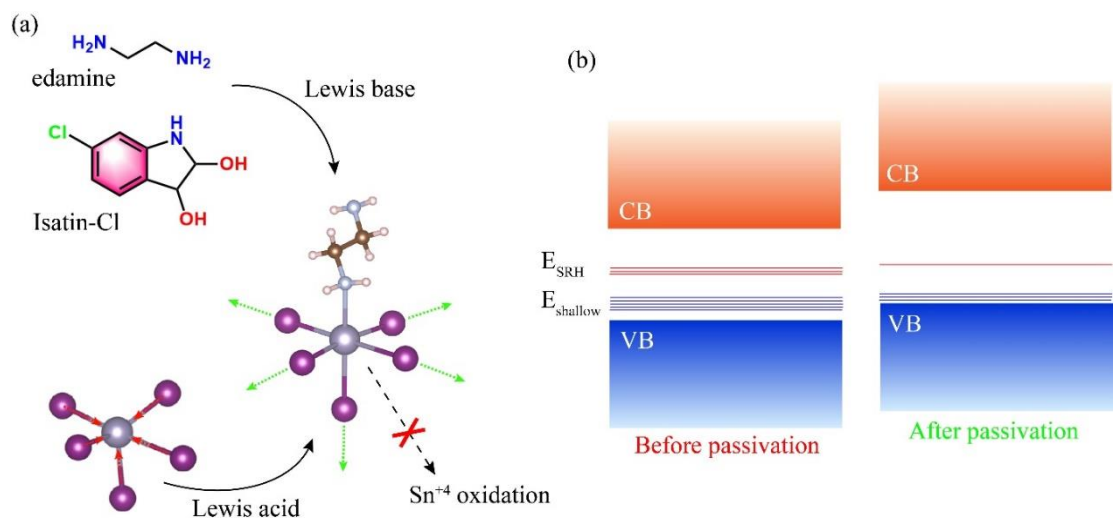

**Figure 7.** Passivation mechanism of Lewis base on tin perovskite surface. (a) schematic showing the molecules induced charge compensation (b) alternating the defect levels before and after passivation.

The band alignment is another critical parameter for interface charge extractions in PSCs, where a reasonable band alignment can reduce the charge transport barrier and enhance the device $V_{oc}$.[22, 54] Aside from the surface passivation, the adsorption of Lewis base molecules can also tune the band alignment of the perovskite layer to match with other charge transfer layers. The effects of the Lewis base molecules on the electronic structure can be dual: firstly, reducing the defective states within the band gap; Secondly, modulating the position of the Fermi level or the work function. Owing to the creation of a dipole layer associated with the molecules' decoration, the work function of the surface can be lowered or increased depending on electron-donating or electron-accepting molecules being adsorbed. The band alignment for the different passivated systems is listed in Fig. S3 including bands of the normal electron and hole transport layers like [6,6]-phenyl-C61-butyric acid methyl ester (PCBM), $TiO_2$-B, and spiro-OMeTAD (Spiro). Baring mind that our pristine surface model contains a high density of unsaturated Sn atoms, such dangling states extend within the band gap of the bulk $FASnI_3$ which makes a very narrow band gap. All the Lewis base molecules are found to passivate such dangling states and broaden the band gap. The upwardly shifted band position proves the released p-type doping by the Lewis base edamine and *p*-TA molecule passivation. For the edamine passivation, the appropriate band position is suitable for acting as the intrinsic layer to connect the electron transfer layer and hole transfer layer in PSCs. In contrast, the significantly lowered band position by the passivation of Isatin-Cl can be used to connect with the hole transfer layer. For the BSA and EDT cases, the rich residual defective states imply an unfavourable photoelectric utilization. Hence, surface molecule adsorption is also an effective method for tuning the band alignment of tin perovskite with other selective layers.

## 4. Conclusion

In summary, we have performed DFT calculations to investigate the passivation ability of soft and hard Lewis base molecules decorating on the defects-contained FASnI$_3$ surface. Our results show that: (1) Lewis base molecules impact the perovskite surface by compensating the background charge and reducing the defect states; (2) the Lewis base molecules mainly interact with the surface uncoordinated tin, inducing charge transfer which compensates the background hole density and avoids further tin oxidation; (3)the healing of defect states leads to an enlarged band gap and less localized V$_{Sn}$ defect; (4) meanwhile, molecular hardness is an important quantity which strongly affects the degree of charge compensation; (5) the hard Lewis base edamine and Isatin-Cl show better stabilizing and passivation function to the hard acid defects-contained tin perovskite surface. (6) Besides, the Lewis base molecular adsorption can effectively regulate the band alignment of tin perovskite. This work provides a basic guideline for the functional Lewis base selection and designing, which is crucial for the fabrication of stable and high-efficiency tin-based PSCs. We hope our work would stimulate future experiments for surface defects passivation via Lewis base molecules. Future studies would be performed with respect to the dynamic electronic excitations in these Lewis base molecular decorated perovskite systems.

**Declaration of Competing Interest**

The authors declare that they have no known competing financial interests or personal relationships that could have appeared to influence the work reported in this paper.

**Acknowledgments**

Funding: This work is supported by the Natural Science Foundation of China [Grant 22022309]; and the Natural Science Foundation of Guangdong Province, China [2021A1515010024], the University of Macau [MYRG2020-00075-IAPME; SRG2019-00179-IAPME], and the Science and Technology Development Fund from Macau SAR [FDCT-0163/2019/A3]. This work was performed in part at the High Performance Computing Cluster (HPCC) which is supported by Information and Communication Technology Office (ICTO) of the University of Macau.